\documentstyle[12pt,axodraw]{article}
\begin{document}
\begin{flushright}
BIHEP-TH-2002-50
\end{flushright}

\renewcommand{\thefootnote}{\fnsymbol{footnote}}

\newcommand{\Title}[1]{{\baselineskip=26pt \begin{center}
            \Large   \bf #1 \\ \ \\ \end{center}}}
 \newcommand{\Author}{\begin{center}\large
      Medina Ablikim\footnote[1]{e-mail: mablikim@mail.ihep.ac.cn},
       Dong-Sheng Du\footnote[2]{e-mail: duds@mail.ihep.ac.cn},
       Mao-Zhi Yang\footnote[3] {e-mail: yangmz@mail.ihep.ac.cn} \end{center}}
\newcommand{\Address}{\begin{center} \it
   Institute of High Energy Physics, Chinese Academy of
             Sciences, P.O.Box 918(4),\\ Beijing 100039, China
\end{center}}

\baselineskip=20pt

\Title{Final State Interactions in $D \to PP$ decays} \Author
\Address
 \vspace{2cm}

\begin{abstract}
The two-body nonleptonic charmed meson decays into two
pseudoscalar mesons are studied using one-particle-exchange
method.  The effects of the final state interactions are analyzed
through the strong phases extracted from the experimental data.
\end{abstract}

\section{Introduction}

The study of the two-body nonleptonic weak decays of particles
containing heavy $(c, b)$ quarks appears to offer a unique
opportunity to determine the basic parameters of quark mixing, and
to investigate the mechanism of $CP$ violation. However, the
quarks are not free, they are bound in hadrons by strong
interactions which are described by nonperturbative QCD. Solving
the problem of nonperturbative QCD needs efforts in both
experiment and theory. In the near future BESIII and CLEO-c
detector will provide high precision data in charm physics
including data on $D$ meson decays, which will provide the
possibility for understanding the physics in charm sector.

It is interesting to study the weak decays of charmed mesons
beyond the factorization approach~\cite{BSW}. In general, if a
process happens in an energy scale where there are many resonance
states, this process must be seriously affected by these
resonances \cite{fsi}. This is a highly nonperturbtive effect.
Near the scale of $D$ meson mass many resonance states exist. $D$
meson decays must be affected seriously by these resonances .
After weak decays the final state particles rescatter into other
particle states through nonperturbative strong interactions
\cite{fsi,Donoghue}.  Every $D$ decay channel can contribute to
each other through final state interaction (FSI). One can model
this rescattering effect as one-particle exchange process
\cite{ope1,ope2}, namely the final state particles be scattered
into other particle states by exchanging one resonance state
existing near the mass scale of $D$ meson. There are also other
ways to treat the nonperturbative and FSI effects in nonleptonic
$D$ decays. One approach is that in which the FSIs are expressed
by the phase shifts of the decaying amplitudes~\cite{ref}. The
other method is a flavor topology approach~\cite{r,cheng}, where
the relative phases between various quark-diagram~\cite{qd}
amplitudes arise from the final state rescattering.

The final state rescattering effects for charmed meson decays into
two pions are studied using the one-particle-exchange
method~\cite{ady}, where the magnitudes of hadronic couplings are
extracted from experimental data on the measured branching
fractions of resonances decays. In addition, a strong phase is
introduced for the hadronic coupling which is important for
obtaining the correct branching ratios in these decays. A similar
analysis is applied to $D \to PV$ decays~\cite{ldy}, where $P$ is
a pseudoscalar meson, and $V$ is vector meson.

In the present work, we extend  the study of the final state
interactions in $D\to \pi\pi$ decays to $D\to PP$ decays. The
coupling constants extracted from experimental data are small for
$s$-channel contribution and large for $t$-channel contribution.
Therefore the $s$-channel contribution is numerically negligible
in $D\to PP$ decays. We safely drop the $s$-channel contribution
in our discussion. In Sec. 2, we present the calculation within
the naive factorization approach. The main scheme of
one-particle-exchange method is described in Sec. 3. We give the
numerical calculations and discussions in Sec. 4. The final
section is reserved for summary.

\section{Calculations in the factorization approach }
\setcounter{equation}{0}

The charmed meson decay can be described by the low energy
effective Hamiltonian\cite{buras}
\begin{equation}
  {\cal H} = \frac{G_F}{\sqrt2} \left[ \sum_{q=d,s} v_q \left(C_1Q_1^q +
         C_2Q_2^q \right)\right], \label{Hamiltonian}
\end{equation}
where $C_1$ and $C_2$ are the Wilson coefficients at $m_c$ scale,
$v_q$ is the product of Cabibbo-Kobayashi-Maskawa (CKM) matrix elements and
defined as
\begin{equation}
  v_q=V_{uq}V_{cq}^* ,
\end{equation}
and the current-current operators are given by
\begin{equation}
  Q_1^q=(\bar u q)_{V-A}(\bar q c)_{V-A}, \qquad
  Q_2^q=(\bar u_\alpha q_\beta)_{V-A}(\bar q_\beta c_\alpha)_{V-A}.
\end{equation}
We do not consider the contributions of QCD and electroweak
penguin operators in the decays of $D \to PP$ because their
contributions are negligible in $D$ decays. QCD factorization
approach~\cite{BBNS} is inapplicable to these decay modes, as the
charmed meson is not heavy enough. The values of $C_1$ and $C_2$
at $m_c$ scale are taken to be \cite{buras}
$$ C_1=1.216,~~~~~C_2=-0.415 .$$

In the naive factorization approach, the decay amplitude can be
generally factorized into a product of two current matrix elements
and can be obtained from~(\ref{Hamiltonian})
\begin{eqnarray}
 && A(D^+ \to \pi^+\pi^0)=-\frac{G_F}{2}\;V_{ud}\;V_{cd}^*\;
      (a_1+a_2)\;if_{\pi}( m_D^2-m_{\pi}^2)\;F^{D\pi}(m_{\pi}^2) ,\nonumber\\
 && A(D^0 \to \pi^+\pi^-)=\frac{G_F}{\sqrt2} \; V_{ud}\;V_{cd}^*\;
      a_1\;if_{\pi}(m_D^2-m_{\pi}^2)\;F^{D\pi}(m_{\pi}^2) ,\nonumber\\
 && A(D^0 \to \pi^0\pi^0)=-\frac{G_F}{2} \; V_{ud}\;V_{cd}^* \;
      a_2\;if_{\pi}(m_D^2-m_{\pi}^2)\;F^{D\pi}(m_{\pi}^2),\nonumber\\
&& A(D^+ \to \bar K^0\pi^+)=\frac{G_F}{\sqrt2} \; V_{ud}
\;V_{cs}^*\left [a_1\;if_{\pi}(m_D^2-m_{K}^2)\;F^{DK}(m_{\pi}^2) \right. \nonumber\\
      & & \hspace*{5.7cm} +\left. a_2\;if_{K}(m_D^2-m_{\pi}^2)\;F^{D\pi}(m_{K}^2) \right], \nonumber\\
&& A(D^0 \to K^-\pi^+)=\frac{G_F}{\sqrt 2} \; V_{ud}\;V_{cs}^* \;
       a_1\;if_{\pi}(m_D^2-m_{K}^2)\;F^{DK}(m_{\pi}^2), \nonumber\\
&& A(D^0 \to \bar K^0\pi^0)=\frac{G_F}{2} \; V_{ud}\;V_{cs}^* \;
       a_2\;if_{K}(m_D^2-m_{\pi}^2)\;F^{D\pi}(m_{K}^2), \nonumber\\
&& A(D^0 \to K^+\pi^-)=\frac{G_F}{\sqrt 2} \; V_{us}\;V_{cd}^* \;
       a_1\;if_{K}(m_D^2-m_{\pi}^2)\;F^{D\pi}(m_{K}^2), \nonumber\\
&& A(D^+ \to K^+\pi^0)=-\frac{G_F}{2} \; V_{us}\;V_{cd}^* \;
       a_1\;if_{K}(m_D^2-m_{\pi}^2)\;F^{D\pi}(m_{K}^2), \nonumber\\
&& A(D^+ \to K^0\pi^+)=\frac{G_F}{\sqrt 2} \; V_{us}\;V_{cd}^* \;
       a_2\;if_{K}(m_D^2-m_{\pi}^2)\;F^{D\pi}(m_{K}^2), \nonumber\\
&& A(D^0 \to K^0\pi^0)=\frac{G_F}{2} \; V_{us}\;V_{cd}^* \;
       a_2\;if_{K}(m_D^2-m_{\pi}^2)\;F^{D\pi}(m_{K}^2),\nonumber\\
&& A(D^+ \to K^+ \bar K^0)=\frac{G_F}{\sqrt 2} \; V_{us}\;V_{cs}^*
        \; a_1 \;if_{K}(m_D^2-m_{K}^2)\;F^{DK}(m_{K}^2), \nonumber\\
&& A(D^0 \to K^+K^-)=\frac{G_F}{\sqrt 2} \; V_{us}\;V_{cs}^* \;
       a_1\;if_{K}(m_D^2-m_{K}^2)\;F^{DK}(m_{K}^2), \nonumber\\
&& A(D^0 \to K^0\bar K^0)=0,
\end{eqnarray}
where the parameters $a_1$ and $a_2$ are defined as~\cite{cheng}
\begin{equation}
a_1=C_1+C_2\left(\frac{1}{N_c}+\chi \right), \qquad
a_2=C_2+C_1\left( \frac{1}{N_c}+\chi\right),
\end{equation}
with the color number $N_c=3$, and $\chi$ is the phenomenological
parameter which takes into account nonfactorizable correction. For
$q^2$ dependence of the form factors, we take the BSW model
\cite{BSW}, i.e., the monopole dominance assumption:
\begin{equation}
F(q^2)=\frac{F(0)}{1-q^2/m_*^2},
\end{equation}
where $m_*$ is the relevant pole mass.

The decay width of a $D$ meson at rest decaying into $PP$ is
\begin{equation}
 \Gamma(D \to PP)=\frac{1}{8\pi}|A(D \to PP)|^2\frac{|\vec
p\;|}{m_D^2},
\end{equation}
where $|\vec p\;|$ is the 3-momentum of each final meson.
The corresponding branching ratio is
\begin{equation}
 Br(D \to PP)=\frac{\Gamma(D \to PP)}{\Gamma_{tot}}.
\end{equation}
\begin{table}[h]
\caption{{\small The branching ratios of $D \to PP$ obtained in
the naive factorization approach and compared with the
experimental results.}}
\begin{center}
\begin{tabular}{|c|c|c|c|} \hline

Decay mode & Br (Theory)  & Br (Theory) & Br(Experiment)\\
           &   $\chi=0$   & $\chi=-8.6 \times 10^{-2}$ &\\
\hline
$D^+ \to \pi^+\pi^0$ &$3.1 \times 10^{-3}$ & $2.71 \times 10^{-3}$ & $(2.5\pm 0.7)\times 10^{-3}$ \\
$D^0 \to \pi^+\pi^-$ &$2.48 \times 10^{-3}$ & $2.65 \times 10^{-3}$&$(1.43 \pm 0.07) \times 10^{-3}$\\
$D^0 \to \pi^0\pi^0$ &$9.98 \times 10^{-8}$  & $1.39 \times 10^{-5}$& $(8.4 \pm 2.2) \times 10^{-4}$\\

$D^+ \to \bar K^0 \pi^+$ &$1.20 \times 10^{-1}$ &$9.98 \times 10^{-2}$ & $(2.77 \pm 0.18)\times 10^{-2}$ \\
$D^0 \to K^- \pi^+$ &$4.81 \times 10^{-2}$ & $5.14 \times 10^{-2}$& $(3.80 \pm 0.09)\times 10^{-2}$ \\
$D^0 \to \bar K^0 \pi^0$ &$2.93 \times 10^{-6}$ & $4.1 \times 10^{-4}$& $(2.28 \pm 0.22)\times 10^{-2}$ \\

$D^0 \to K^+ \pi^-$ &$1.88 \times 10^{-4}$ & $2.01 \times 10^{-4}$& $(1.48 \pm 0.21)\times 10^{-4}$ \\
$D^+ \to K^+ \pi^0$ &$2.4 \times 10^{-4}$ &$2.56 \times 10^{-4}$ & - \\
$D^+ \to K^0 \pi^+$ &$3.86 \times 10^{-8}$ & $5.39 \times 10^{-6}$& - \\
$D^0 \to K^0 \pi^0$ &$7.58 \times 10^{-9}$ & $1.06 \times 10^{-6}$& - \\

$D^0 \to K^+ \bar K^0$ &$9.15 \times 10^{-3}$ &$9.76 \times 10^{-3}$ & $(5.8 \pm 0.6) \times 10^{-3}$\\
$D^0 \to K^+ K^-$ &$3.59\times 10^{-3}$  & $3.83 \times 10^{-3}$& $(4.12 \pm 0.14) \times 10^{-3}$\\
$D^0 \to K^0 \bar K^0$ & 0 & 0 & $(7.1 \pm 1.9) \times 10^{-4}$\\
\hline
\end{tabular}
\end{center}
\end{table}
A comparison of the branching ratios of the naive factorization
result with the experimental data is presented in Table 1. The
second column gives the pure factorization result, where the
nonfactorization effect  is zero, while the third column
represents the branching ratio with small nonfactorization
correction. One can notice that the results are not in agreement
with the experimental data. For doubly Cabibbo-suppressed decay
modes, the experimental measurements of their decay rates are unavailable,
except for the channel $D^0 \to K^+ \pi^-$. We shall predict their
theoretical branching ratios in section 4. The ratio for
Cabibbo-suppressed decay mode $D^0 \to K^0 \bar K^0$ vanishes in
the naive factorization approach. This decay seems to be induced
through final state rescattering.

\section{The one particle exchange method for FSI}

As we have seen above, the experimental results for the branching
ratios are mostly in disagreement with the calculation from the
naive factorization approach. The reason is that the physical
picture of naive factorization is too simple, nonperturbative
strong interactions are restricted in single hadrons, or between
the initial and final hadrons which share the same spectator
quark. If the mass of the initial particle is large, such as the
case of $B$ meson decay, the effect of nonperturbative strong
interactions between the final hadrons most probably is small
because the momentum transfer is large. However, in the case of
$D$ meson, its mass is not so large. The energy scale of $D$
decays is not very high. Nonperturbative effects may give large
contribution. Because there exist many resonances near the mass
scale of $D$ meson, it is possible that nonperturbative
interactions propagate through these resonance states, such as,
$K^*(892)$, $K^*(1430)$, $f_0(1710)$, $\rho(770)$, $\phi(1020)$
etc.

\begin{figure}[h]
\begin{center}
\begin{picture}(250,90)
      \put(35,50){\line(1,0){30}}
      \put(69,50){\circle*{8}}
      \put(87,50){\circle{80}}
      \put(108,50){\circle*{8}}
      \put(112,50){\line(1,0){38}}
      \put(148,50){\circle*{8}}
      \put(148,50){\line(3,-2){45}}
      \put(148,50){\line(3,2){45}}

   \put(50,50){\vector(3,0){2}}
   \put(88,70){\vector(3,0){2}}
   \put(88,30){\vector(3,0){2}}
   \put(130,50){\vector(3,0){2}}
   \put(170,64.5){\vector(3,2){2}}
   \put(170,35.5){\vector(3,-2){2}}

      \put(10,48){$D$}
      \put(80,16){$P_2$}
      \put(80,76){$P_1$}
      \put(122,59){$0^{++}$}
      \put(200,78){$P_3$}
      \put(200,13){$P_4$}
   \end{picture}
 \end{center}
 \caption{{\small s-channel contributions to final-state interactions in $D\to
     PP $ due to one particle  exchange.}}
 \label{schan}
\end{figure}
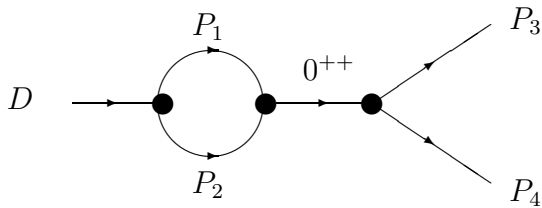

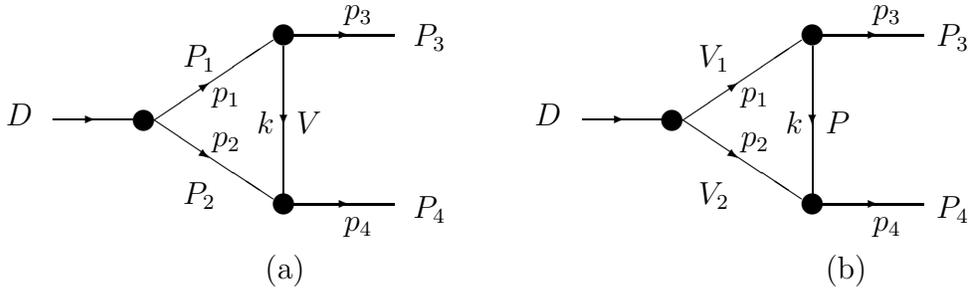
\begin{figure}[h]
\begin{center}
\begin{picture}(430,160)
\put(28,65){\line(1,0){30}}
\put(62,65){\circle*{8}}
\put(66,65){\line(3,2) {45}}
\put(66,65){\line(3,-2){45}}
\put(115,33){\circle*{8}}
\put(115,97){\circle*{8}}
\put(115,37){\line(0,1){56}}
\put(119,33){\line(1,0){38}}
\put(119,97){\line(1,0){38}}

\put(42,65){\vector(3,0){2}}
\put(85,52){\vector(3,-2){2}}
\put(85,77.5){\vector(3,2){2}}
\put(115,65){\vector(0,-2){2}}
\put(138,97){\vector(3,0){2}}
\put(138,33){\vector(3,0){2}}

\put(10,63){$D$}
\put(77,33){$P_2$}
\put(77,85){$P_1$}
\put(120,60){$V$}
\put(88,72){$p_1$}
\put(88,55){$p_2$}
\put(105,60){$k$}
\put(164,93){$P_3$}
\put(164,28){$P_4$}
\put(138,103){$p_3$}
\put(138,23){$p_4$}
\put(108,5){(a)}

\put(228,65){\line(1,0){30}}
\put(262,65){\circle*{8}}
\put(266,65){\line(3,2) {45}}
\put(266,65){\line(3,-2){45}}
\put(315,33){\circle*{8}}
\put(315,97){\circle*{8}}
\put(315,37){\line(0,1){56}}
\put(319,33){\line(1,0){38}}
\put(319,97){\line(1,0){38}}

 \put(242,65){\vector(3,0){2}}
 \put(285,52){\vector(3,-2){2}}
 \put(285,77.5){\vector(3,2){2}}
 \put(315,65){\vector(0,-2){2}}
 \put(338,97){\vector(3,0){2}}
 \put(338,33){\vector(3,0){2}}

  \put(210,63){$D$}
  \put(272,33){$V_2$}
  \put(272,85){$V_1$}
  \put(320,60){$P$}
  \put(288,72){$p_1$}
  \put(288,55){$p_2$}
  \put(305,60){$k$}
  \put(362,93){$P_3$}
  \put(362,28){$P_4$}
  \put(338,103){$p_3$}
  \put(338,23){$p_4$}
  \put(320,5){(b)}
\end{picture}
\end{center}
\caption{{\small t-channel contributions to final-state
interactions in $D\to PP $ due to one particle  exchange.
 (a) Exchange a single vector meson, (b) Exchange a single
pseudoscalar meson}}
 \label{tchan}
\end{figure}

The diagrams of these nonperturbative rescattering effects can be
depicted in Figs.\ref{schan} and \ref{tchan}. The first part $D
\to P_1P_2$ or $D \to V_1V_2$ represents the direct decay where
the decay amplitudes can be obtained by using naive factorization
method. The second part represents rescattering process where the
effective hadronic couplings are needed in numerical calculation,
which can be extracted from experimental data on the relevant
resonance decays.

Fig.\ref{schan} is the $s$-channel contributions to the final
state interactions. Here $P_1$ and $P_2$ are the intermediate
pseudoscalar mesons. The resonance state has the quantum number
$J^{PC}=0^{++}$ derived from the final state particles $P_3$ and
$P_4$. From Particle Data Group~\cite{PDG}, one can only choose
$f_0(1710)$ as the resonance state which evaluates the $s$-channel
contribution. However, the coupling of $f_0(1710)$ with two final
mesons $P_3$ and $P_4$ is too small~\cite{ady}, we drop the
$s$-channel contribution in the numerical calculation.

Fig.\ref{tchan} shows the $t$-channel contribution to the final
state interactions. $P_1$, $P_2$ and $V_1$, $V_2$ are the
intermediate states. They rescatter into the final state
$P_{3}P_{4}$ by exchanging one resonance state $V$ or $P$. In this
paper the intermediate states are treated to be on their mass
shell, because their off-shell contribution can be attributed to
the quark level. We assume the on-shell contribution dominates in
the final state interactions. The exchanged resonances are treated
as a virtual particle. Their propagators are taken as Breit-Wigner
form
\begin{equation}
\frac{i}{k^2-m^2+im\Gamma_{tot}}, \nonumber
\end{equation}
where $\Gamma_{tot}$ is the total decay width of the exchanged resonance.
To the lowest order, the effective couplings of $f_0$ to $PP$ and
$VV$ can be taken as the form
\begin{eqnarray}
L_I & =& g_{fPP} \phi^+\phi f,\\
L_I & =& g_{fVV} A_\mu A^\mu f,
\end{eqnarray}
where $\phi$ is the pseudoscalar field, $A_\mu$ the vector field.
Then the decay amplitudes of $f_0\to PP$ and $VV$ are
\begin{eqnarray}
T_{fPP} & =& g_{fPP},\label{t1}\\
T_{fVV} & =& g_{fVV} \epsilon_\mu\epsilon^{\mu}.
\end{eqnarray}
The coupling constants $g_{fPP}$ and $g_{fVV}$ can be extracted
from the measured branching fractions of $f_0\to PP$ and $VV$
decays, respectively \cite{PDG}.  Because $f_0\to VV$ decays have
not been detected yet, we assume their couplings are small. We do
not consider the intermediate vector meson contributions of
$s$-channel in this paper.

For the $t$-channel contribution, the concerned effective vertex
is $VPP$, which can be related to the $V$ decay amplitude.
Explicitly the amplitude of $V\to PP$ can be written as
\begin{equation}
T_{VPP}=g_{VPP}\;\epsilon\cdot (p_1-p_2),
\label{t2}
\end{equation}
where $p_1$ and $p_2$ are the four-momentum of the two
pseudoscalars, respectively. To extract $g_{fPP}$ and $g_{VPP}$
from experiment, one should square eqs.(\ref{t1}) and (\ref{t2})
to get the decay widths
\begin{eqnarray}
\Gamma(f\to PP)&=& \frac{1}{8\pi}\mid g_{fPP}\mid^2
           \frac{\mid\vec{p}\mid}{m_f^2},\nonumber\\
\Gamma(V\to PP)&=& \frac{1}{3}\frac{1}{8\pi}\mid g_{VPP}\mid^2
         \left[m_V^2-2m_1^2-2m_2^2+\frac{(m_1^2-m_2^2)^2}{m_V^2}\right]
          \frac{\mid\vec{p}\mid}{m_V^2},
\label{couple}
\end{eqnarray}

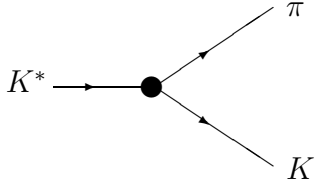
\begin{figure}
\begin{center}
\begin{picture}(150,60)

\put(28,35){\line(1,0){38}} \put(65,35){\circle*{8}}
\put(66,35){\line(3,2) {45}} \put(66,35){\line(3,-2){45}}

\put(42,35){\vector(3,0){2}} \put(85,22){\vector(3,-2){2}}
\put(85,47.5){\vector(3,2){2}}

\put(10,33){$K^*$} \put(116,0){$K$} \put(116,62){$\pi$}

\end{picture}
\end{center}
\caption{{\small The effective coupling vertex on the hadronic
level}} \label{vertex1}
\end{figure}
\begin{figure}
\begin{center}
\begin{picture}(150,100)

\put(36,65){\line(3,0){30}} \put(36,40){\line(3,0){30}}
\put(66,65){\line(3,2) {45}} \put(66,40){\line(3,-2){45}}
\put(73,52){\line(3,2) {45}} \put(73,52){\line(3,-2){45}}

\GlueArc(63,67)(10,195,375){2}{8}
\GlueArc(63,38)(10,-20,170){2}{8} \Gluon(66,65)(73,52){2}{3}
\Gluon(66,40)(73,52){2}{3} \GlueArc(73,60)(15,160,300){2}{15}

\put(42,65){\vector(3,0){2}} \put(42,40){\vector(-3,0){2}}

\put(100,88){\vector(3,2){2}} \put(105,73.5){\vector(-3,-2){2}}

\put(105,30.5){\vector(3,-2){2}} \put(99,18){\vector(-3,2){2}}

\put(-8,50){$K^{*+}$} \put(23,63){$u$} \put(23,38){$\bar s$}
\put(100,4){$\bar s$} \put(100,96){$u$} \put(110,34){$d$}
\put(110,66){$\bar d$}
\put(125,6){$K^0$} \put(128,91){$\pi^+$}

\end{picture}
\end{center}
\caption{{\small The effective coupling vertex on the quark
level}} \label{vertex2}
\end{figure}
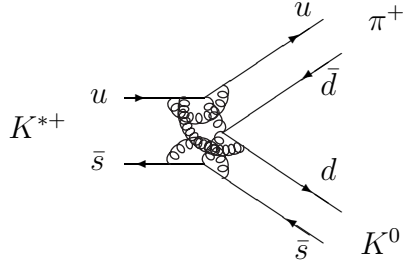

\noindent where $m_1$ and $m_2$ are the masses of the two final
particles $PP$, respectively, and $\mid\vec{p}\mid$ is the
momentum of one of the final particle $P$ in the rest frame of $V$
or $f$. From the above equations, one can see that only the
magnitudes of the effective couplings $\mid g_{fPP}\mid $ and
$\mid g_{VPP}\mid $ can be extracted from experiment. If there is
any phase factor for the effective coupling, it would be dropped.
Actually it is quite possible that there are imaginary phases for
the effective couplings. As an example, let us see the effective
coupling of $g_{K^*K\pi}$ shown in Fig.\ref{vertex1}, which is
relevant to the process $K^* \to K\pi$. On the quark level, the
effective vertex can be depicted in Fig.\ref{vertex2}, which
should be controlled by nonperturbative QCD. From this figure one
can see that it is reasonable that a strong phase could appear in
the effective coupling, which is contributed by strong
interactions. Therefore we can introduce a strong phase for each
hadronic effective coupling. In the succeeding part of this paper,
the symbol $g$ will only be used to represent the magnitude of the
relevant effective coupling. The total one should be $g e^{i
\theta}$, where $\theta$ is the strong phase coming from
Fig.\ref{vertex2}. For example, the effective couplings will be
written in the form of $g_{fPP}e^{i\theta_{fPP}}$ and
$g_{VPP}e^{i\theta_{VPP}}$.

The decay amplitude of the $s$-channel final state interactions
can be calculated from Fig.\ref{schan}
\begin{equation}
  A^{FSI}_s=\frac12 \int \frac{d^3 \vec p_1}{(2\pi)^3
2E_1} \int \frac{d^3 \vec p_2}{(2\pi)^3 2E_2} (2\pi)^4
\delta^4(p_D-p_1-p_2) \; A(D \to
P_1P_2)\frac{i\;g_1\;g_2\;e^{i(\theta_1+\theta_2)}}{k^2-m^2+im\Gamma_{tot}},
\end{equation}
where $p_1$ and $p_2$ represent the four-momenta of the
pseudoscalar $P_1$ and $P_2$, the amplitude $A(D \to P_1P_2)$ is
the direct decay amplitude.  The effective coupling constants
$g_1$ and $g_2$ should be $g_{fPP}$ or $g_{VPP}$ which can be
obtained by comparing eq.(\ref{couple}) with experimental data.
By performing integrals, we obtain
\begin{equation}
  A^{FSI}_s=\frac{1}{8\pi m_D} |\vec p_1| A(D \to P_1P_2)
  \frac{i\;g_1\;g_2\;e^{i(\theta_1+\theta_2)}}{k^2-m^2+im\Gamma_{tot}}.
\end{equation}

The $t$-channel contribution via exchanging a vector meson
(Fig.\ref{tchan}(a) ) is
\begin{eqnarray}
  A^{FSI}_{t,V}&=&\frac12 \int \frac{d^3 \vec p_1}{(2\pi)^3 2E_1}
          \int\frac{d^3 \vec p_2}{(2\pi)^3 2E_2} (2\pi)^4
 \delta^4(p_D-p_1-p_2) A(D \to P_1P_2)\nonumber\\
  & &\hspace*{1cm}\times g_1 \; \epsilon_\lambda \cdot (p_1+p_3)
  \frac{i\;e^{i(\theta_1+\theta_2)}}{k^2-m^2+im\Gamma_{tot}} \;
 F(k^2)^2 \;g_2 \; \epsilon_\lambda\cdot(p_2+p_4),
\label{fsisk}
\end{eqnarray}
where $F(k^2)=(\Lambda^2-m^2)/(\Lambda^2-k^2)$ is the form factor
which is introduced to compensate the off-shell effect of the
exchanged particle at the vertices \cite{Fk}. We choose the
lightest resonance state as the exchanged particle that gives rise
to the largest contribution to the decay amplitude.

We furthermore have
\begin{equation}
  A^{FSI}_{t,V}=\int^1_{-1}  \frac{d(\cos\theta)}{16\pi m_D} |\vec p_1|
   A(D \to P_1P_2) \;g_1 \frac{i\;e^{i(\theta_1+\theta_2)}}
  {k^2-m^2+im\Gamma_{tot}} \;F(k^2)^2\;g_2\;H ,
\label{fsitv}
\end{equation}
where
\begin{eqnarray}
  H&=&-\left[m_D^2 -\frac12(m_1^2+m_2^2+m_3^2+m_4^2)
     +(|\vec p_1||\vec p_4|+ |\vec p_2||\vec p_3|)\cos\theta
     +E_1E_4+E_2E_3 \right]\nonumber\\
   & & -\frac{1}{m_V^2}(m_1^2-m_3^2)(m_2^2-m_4^2).
\end{eqnarray}
The $t$-channel contribution by exchanging a pseudoscalar meson
(Fig.\ref{tchan}(b) )is
\begin{eqnarray}
  A^{FSI}_{t,P}&=&\frac12 \int \frac{d^3 \vec p_1}{(2\pi)^3 2E_1}
          \int\frac{d^3 \vec p_2}{(2\pi)^3 2E_2} (2\pi)^4
 \delta^4(p_D-p_1-p_2) \sum_{\lambda_1,\lambda_2}A(D \to V_1V_2)\nonumber\\
  & &\hspace*{1cm} \times g_1 \; \epsilon_{\lambda_1} \cdot (p_3-k)
  \frac{i\;e^{i(\theta_1+\theta_2)}}{k^2-m^2+im\Gamma_{tot}} \; F(k^2)^2
    \;g_2 \; \epsilon_{\lambda_2}\cdot(k+p_4),
\end{eqnarray}
and we obtain
\begin{eqnarray}
 A^{FSI}_{t,P}=\int^1_{-1} \frac{d(\cos\theta)}{16\pi m_D} |\vec p_1| \;
    \frac{i\;e^{i(\theta_1+\theta_2)}}{k^2-m^2+im\Gamma_{tot}}
      X \; g_1 \;g_2 \; F(k^2)^2\;(-H_1+H_2),
\label{fsitp}
\end{eqnarray}
where
\begin{eqnarray}
  H_1&=&4im_{V_1}f_{V_1}(m_D+m_2)
A_1\left[\frac12(m_D^2-m_3^2-m_4^2)\right.\nonumber\\
    &&\hspace*{4cm} -\frac{1}{m_1^2} ( E_1E_3-|\vec p_1||\vec
p_3|\cos\theta)
          ( E_1E_4+|\vec p_1||\vec p_4|\cos\theta)\nonumber\\
  &&\hspace*{4cm} -\frac{1}{m_2^2}( E_2E_4-|\vec p_2||\vec
      p_4|\cos\theta) ( E_2E_3+|\vec p_2||\vec p_3|\cos\theta)\nonumber\\
& &\left.+\frac{1}{2m_1^2m_2^2}( m_D^2-m_1^2-m_2^2 )
 ( E_1E_3-|\vec p_1||\vec p_3|\cos\theta )(E_2E_4-|\vec p_2||\vec
p_4|\cos\theta) \right],
\end{eqnarray}
\begin{eqnarray}
H_2&=&\frac{8im_{V_1}f_{V_1}}{(m_D+m_2)}A_2
     \left[ E_2E_3+|\vec p_2||\vec p_3|\cos\theta -
  \frac{1}{2m_1^2}(m_D^2-m_1^2-m_2^2)(E_1E_3-|\vec p_1||\vec
p_3|\cos\theta)\right]\nonumber\\
& &\hspace*{1cm}\left[ E_1E_4+|\vec p_1||\vec p_4|\cos\theta -
  \frac{1}{2m_2^2}(m_D^2-m_1^2-m_2^2)( E_2E_4-|\vec p_2||\vec
p_4|\cos\theta ) \right],
\end{eqnarray}
and $X$ represents the relevant direct decay amplitude of $D$
decaying to the intermediate vector pair $V_1$ and $V_2$ divided
by $\langle V_1 |(V-A)_\mu |0\rangle \langle
V_2|(V-A)^\mu|D\rangle$,
$$
X\equiv \frac{A(D\to V_1 V_2)}
  {\langle V_1 |(V-A)_\mu |0\rangle \langle
  V_2|(V-A)^\mu|D\rangle}.
$$

\section{Numerical calculation and  discussion}

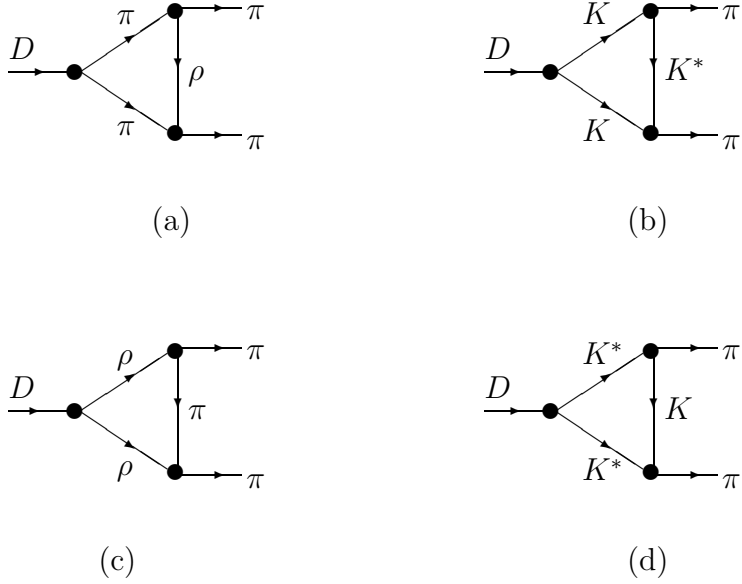
\begin{figure}
\begin{center}
\begin{picture}(480,100)

\put(91,65){\line(1,0){24}} \put(116,65){\circle*{6}}
\put(118,65){\line(3,2) {33}} \put(118,65){\line(3,-2){33}}
\put(154,42){\circle*{6}} \put(154,88){\circle*{6}}
\put(155,44){\line(0,1){44}} \put(155,41){\line(1,0){24}}
\put(155,89){\line(1,0){24}}

\put(103,65){\vector(3,0){2}} \put(137,52){\vector(3,-2){2}}
\put(137,78){\vector(3,2){2}} \put(155,68){\vector(0,-2){2}}
\put(171,89){\vector(3,0){2}} \put(171,41){\vector(3,0){2}}

\put(91,70){$D$} \put(132,41){$\pi$} \put(132,83){$\pi$}
\put(159,62){$\rho$} \put(181,85){$\pi$} \put(181,36){$\pi$}
\put(145,5){(a)}


\put(271,65){\line(1,0){24}} \put(296,65){\circle*{6}}
\put(298,65){\line(3,2) {33}} \put(298,65){\line(3,-2){33}}
\put(334,42){\circle*{6}} \put(334,88){\circle*{6}}
\put(335,44){\line(0,1){44}} \put(335,41){\line(1,0){24}}
\put(335,89){\line(1,0){24}}

\put(283,65){\vector(3,0){2}} \put(317,52){\vector(3,-2){2}}
\put(317,78){\vector(3,2){2}} \put(335,68){\vector(0,-2){2}}
\put(351,89){\vector(3,0){2}} \put(351,41){\vector(3,0){2}}

\put(271,70){$D$} \put(308,39){$K$} \put(308,83){$K$}
\put(339,62){$K^*$} \put(361,85){$\pi$} \put(361,36){$\pi$}
\put(325,5){(b)}

\end{picture}
\end{center}

\begin{center}
\begin{picture}(480,100)

\put(91,65){\line(1,0){24}} \put(116,65){\circle*{6}}
\put(118,65){\line(3,2) {33}} \put(118,65){\line(3,-2){33}}
\put(154,42){\circle*{6}} \put(154,88){\circle*{6}}
\put(155,44){\line(0,1){44}} \put(155,41){\line(1,0){24}}
\put(155,89){\line(1,0){24}}

\put(101,65){\vector(3,0){2}} \put(137,52){\vector(3,-2){2}}
\put(137,78){\vector(3,2){2}} \put(155,68){\vector(0,-2){2}}
\put(171,89){\vector(3,0){2}} \put(171,41){\vector(3,0){2}}

\put(91,70){$D$} \put(132,41){$\rho$} \put(132,83){$\rho$}
\put(159,62){$\pi$} \put(181,85){$\pi$} \put(181,36){$\pi$}
\put(125,5){(c)}

\put(271,65){\line(1,0){24}} \put(296,65){\circle*{6}}
\put(298,65){\line(3,2) {33}} \put(298,65){\line(3,-2){33}}
\put(334,42){\circle*{6}} \put(334,88){\circle*{6}}
\put(335,44){\line(0,1){44}} \put(335,41){\line(1,0){24}}
\put(335,89){\line(1,0){24}}

\put(283,65){\vector(3,0){2}} \put(317,52){\vector(3,-2){2}}
\put(317,78){\vector(3,2){2}} \put(335,68){\vector(0,-2){2}}
\put(351,89){\vector(3,0){2}} \put(351,41){\vector(3,0){2}}

\put(271,70){$D$} \put(308,39){$K^*$} \put(308,83){$K^*$}
\put(339,62){$K$} \put(361,85){$\pi$} \put(361,36){$\pi$}
\put(325,5){(d)}

\end{picture}
\end{center}

\caption{{\small Intermediate states in rescattering process for
$D\to\pi\pi$ decays}} \label{scatter1}
\end{figure}

\begin{figure}
\begin{center}
\begin{picture}(480,100)
\put(36,65){\line(1,0){24}} \put(61,65){\circle*{6}}
\put(63,65){\line(3,2) {33}} \put(63,65){\line(3,-2){33}}
\put(99,42){\circle*{6}} \put(99,88){\circle*{6}}
\put(100,44){\line(0,1){44}} \put(100,41){\line(1,0){24}}
\put(100,89){\line(1,0){24}}

\put(48,65){\vector(3,0){2}} \put(82,52){\vector(3,-2){2}}
\put(82,78){\vector(3,2){2}} \put(100,68){\vector(0,-2){2}}
\put(116,89){\vector(3,0){2}} \put(116,41){\vector(3,0){2}}

\put(36,70){$D$} \put(77,38){$K$} \put(77,83){$\pi$}
\put(104,62){$K^*$} \put(126,85){$K$} \put(126,36){$\pi$}
\put(90,5){(a)}


\put(186,65){\line(1,0){24}} \put(211,65){\circle*{6}}
\put(213,65){\line(3,2) {33}} \put(213,65){\line(3,-2){33}}
\put(249,42){\circle*{6}} \put(249,88){\circle*{6}}
\put(250,44){\line(0,1){44}} \put(250,41){\line(1,0){24}}
\put(250,89){\line(1,0){24}}

\put(197,65){\vector(3,0){2}} \put(232,52){\vector(3,-2){2}}
\put(232,78){\vector(3,2){2}} \put(250,68){\vector(0,-2){2}}
\put(266,89){\vector(3,0){2}} \put(266,41){\vector(3,0){2}}

\put(186,70){$D$} \put(227,38){$K$} \put(227,83){$\pi$}
\put(254,62){$\rho$} \put(276,85){$\pi$} \put(276,36){$K$}
\put(240,5){(b)}

\put(336,65){\line(1,0){24}} \put(361,65){\circle*{6}}
\put(363,65){\line(3,2) {33}} \put(363,65){\line(3,-2){33}}
\put(399,42){\circle*{6}} \put(399,88){\circle*{6}}
\put(400,44){\line(0,1){44}} \put(400,41){\line(1,0){24}}
\put(400,89){\line(1,0){24}}

\put(348,65){\vector(3,0){2}} \put(382,52){\vector(3,-2){2}}
\put(382,78){\vector(3,2){2}} \put(400,68){\vector(0,-2){2}}
\put(416,89){\vector(3,0){2}} \put(416,41){\vector(3,0){2}}

\put(336,70){$D$} \put(377,38){$K$} \put(377,83){$\pi$}
\put(404,62){$K^*$} \put(426,85){$\pi$} \put(426,36){$K$}
\put(390,5){(c)}

\end{picture}
\end{center}
\begin{center}
\begin{picture}(480,100)

\put(36,65){\line(1,0){24}} \put(61,65){\circle*{6}}
\put(63,65){\line(3,2) {33}} \put(63,65){\line(3,-2){33}}
\put(99,42){\circle*{6}} \put(99,88){\circle*{6}}
\put(100,44){\line(0,1){44}} \put(100,41){\line(1,0){24}}
\put(100,89){\line(1,0){24}}

\put(46,65){\vector(3,0){2}} \put(82,52){\vector(3,-2){2}}
\put(82,78){\vector(3,2){2}} \put(100,68){\vector(0,-2){2}}
\put(116,89){\vector(3,0){2}} \put(116,41){\vector(3,0){2}}

\put(36,70){$D$} \put(76,38){$K^*$} \put(76,83){$\rho$}
\put(104,62){$K$} \put(126,85){$K$} \put(126,36){$\pi$}
\put(70,5){(d)}


\put(186,65){\line(1,0){24}} \put(211,65){\circle*{6}}
\put(213,65){\line(3,2) {33}} \put(213,65){\line(3,-2){33}}
\put(249,42){\circle*{6}} \put(249,88){\circle*{6}}
\put(250,44){\line(0,1){44}} \put(250,41){\line(1,0){24}}
\put(250,89){\line(1,0){24}}

\put(197,65){\vector(3,0){2}} \put(232,52){\vector(3,-2){2}}
\put(232,78){\vector(3,2){2}} \put(250,68){\vector(0,-2){2}}
\put(266,89){\vector(3,0){2}} \put(266,41){\vector(3,0){2}}

\put(186,70){$D$} \put(225,38){$K^*$} \put(225,83){$\rho$}
\put(254,62){$\pi$} \put(276,85){$\pi$} \put(276,36){$K$}
\put(240,5){(e)}

\put(336,65){\line(1,0){24}} \put(361,65){\circle*{6}}
\put(363,65){\line(3,2) {33}} \put(363,65){\line(3,-2){33}}
\put(399,42){\circle*{6}} \put(399,88){\circle*{6}}
\put(400,44){\line(0,1){44}} \put(400,41){\line(1,0){24}}
\put(400,89){\line(1,0){24}}

\put(348,65){\vector(3,0){2}} \put(382,52){\vector(3,-2){2}}
\put(382,78){\vector(3,2){2}} \put(400,68){\vector(0,-2){2}}
\put(416,89){\vector(3,0){2}} \put(416,41){\vector(3,0){2}}

\put(336,70){$D$} \put(376,38){$K^*$} \put(376,83){$\rho$}
\put(404,62){$K$} \put(426,85){$\pi$} \put(426,36){$K$}
\put(390,5){(f)}
\end{picture}
\end{center}

\caption{{\small Intermediate states in rescattering process for
$D\to K\pi$ decays}} \label{scatter2}
\end{figure}
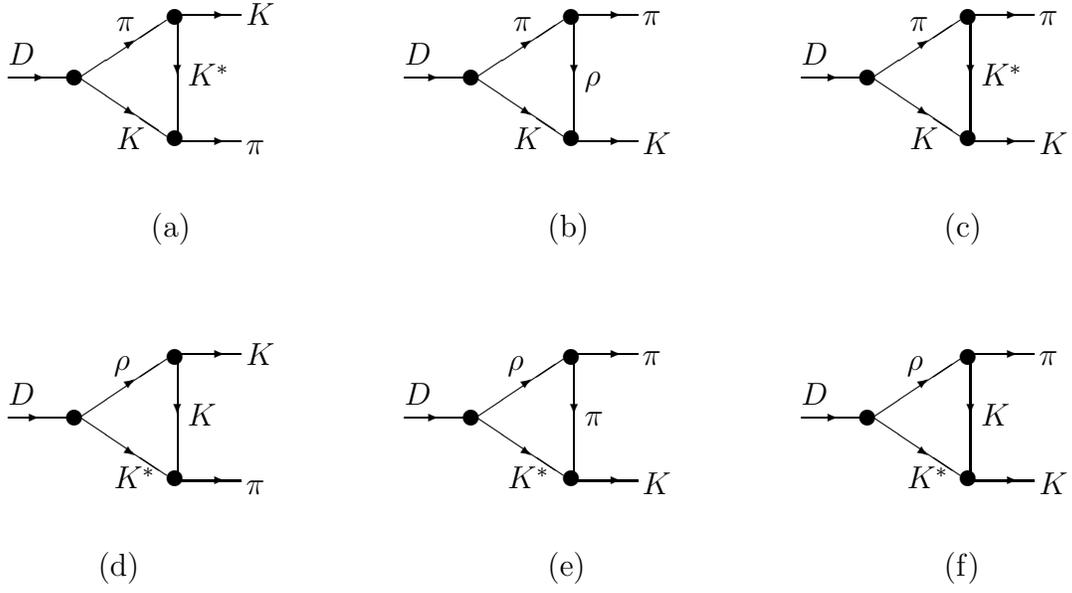

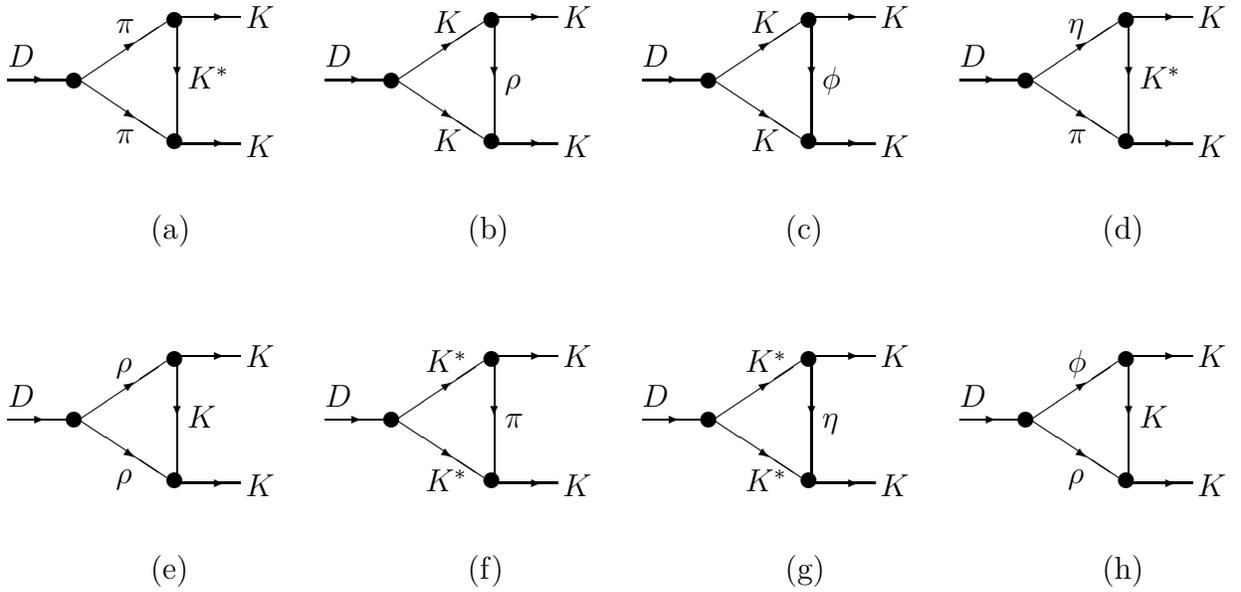
\begin{figure}
\begin{center}
\begin{picture}(480,100)
\put(6,65){\line(1,0){24}} \put(31,65){\circle*{6}}
\put(33,65){\line(3,2) {33}} \put(33,65){\line(3,-2){33}}
\put(69,42){\circle*{6}} \put(69,88){\circle*{6}}
\put(70,44){\line(0,1){44}} \put(70,41){\line(1,0){24}}
\put(70,89){\line(1,0){24}}

\put(18,65){\vector(3,0){2}} \put(52,52){\vector(3,-2){2}}
\put(52,78){\vector(3,2){2}} \put(70,68){\vector(0,-2){2}}
\put(86,89){\vector(3,0){2}} \put(86,41){\vector(3,0){2}}

\put(6,70){$D$} \put(47,41){$\pi$} \put(47,83){$\pi$}
\put(74,62){$K^*$} \put(96,85){$K$} \put(96,36){$K$}
\put(60,5){(a)}

\put(126,65){\line(1,0){24}} \put(151,65){\circle*{6}}
\put(153,65){\line(3,2) {33}} \put(153,65){\line(3,-2){33}}
\put(189,42){\circle*{6}} \put(189,88){\circle*{6}}
\put(190,44){\line(0,1){44}} \put(190,41){\line(1,0){24}}
\put(190,89){\line(1,0){24}}

\put(138,65){\vector(3,0){2}} \put(172,52){\vector(3,-2){2}}
\put(172,78){\vector(3,2){2}} \put(190,68){\vector(0,-2){2}}
\put(206,89){\vector(3,0){2}} \put(206,41){\vector(3,0){2}}

\put(126,70){$D$} \put(167,38){$K$} \put(167,83){$K$}
\put(194,62){$\rho$} \put(216,85){$K$} \put(216,36){$K$}
\put(180,5){(b)}


\put(246,65){\line(1,0){24}} \put(271,65){\circle*{6}}
\put(273,65){\line(3,2) {33}} \put(273,65){\line(3,-2){33}}
\put(309,42){\circle*{6}} \put(309,88){\circle*{6}}
\put(310,44){\line(0,1){44}} \put(310,41){\line(1,0){24}}
\put(310,89){\line(1,0){24}}

\put(258,65){\vector(3,0){2}} \put(292,52){\vector(3,-2){2}}
\put(292,78){\vector(3,2){2}} \put(310,68){\vector(0,-2){2}}
\put(326,89){\vector(3,0){2}} \put(326,41){\vector(3,0){2}}

\put(246,70){$D$} \put(287,38){$K$} \put(287,83){$K$}
\put(314,62){$\phi$} \put(336,85){$K$} \put(336,36){$K$}
\put(300,5){(c)}

\put(366,65){\line(1,0){24}} \put(391,65){\circle*{6}}
\put(393,65){\line(3,2) {33}} \put(393,65){\line(3,-2){33}}
\put(429,42){\circle*{6}} \put(429,88){\circle*{6}}
\put(430,44){\line(0,1){44}} \put(430,41){\line(1,0){24}}
\put(430,89){\line(1,0){24}}

\put(378,65){\vector(3,0){2}} \put(412,52){\vector(3,-2){2}}
\put(412,78){\vector(3,2){2}} \put(430,68){\vector(0,-2){2}}
\put(446,89){\vector(3,0){2}} \put(446,41){\vector(3,0){2}}

\put(366,70){$D$} \put(407,41){$\pi$} \put(407,83){$\eta$}
\put(434,62){$K^*$} \put(456,85){$K$} \put(456,36){$K$}
\put(420,5){(d)}

\end{picture}
\end{center}
\begin{center}
\begin{picture}(480,100)

\put(6,65){\line(1,0){24}} \put(31,65){\circle*{6}}
\put(33,65){\line(3,2) {33}} \put(33,65){\line(3,-2){33}}
\put(69,42){\circle*{6}} \put(69,88){\circle*{6}}
\put(70,44){\line(0,1){44}} \put(70,41){\line(1,0){24}}
\put(70,89){\line(1,0){24}}

\put(18,65){\vector(3,0){2}} \put(52,52){\vector(3,-2){2}}
\put(52,78){\vector(3,2){2}} \put(70,68){\vector(0,-2){2}}
\put(86,89){\vector(3,0){2}} \put(86,41){\vector(3,0){2}}

\put(6,70){$D$} \put(47,41){$\rho$} \put(47,83){$\rho$}
\put(74,62){$K$} \put(96,85){$K$} \put(96,36){$K$} \put(60,5){(e)}

\put(126,65){\line(1,0){24}} \put(151,65){\circle*{6}}
\put(153,65){\line(3,2) {33}} \put(153,65){\line(3,-2){33}}
\put(189,42){\circle*{6}} \put(189,88){\circle*{6}}
\put(190,44){\line(0,1){44}} \put(190,41){\line(1,0){24}}
\put(190,89){\line(1,0){24}}

\put(138,65){\vector(3,0){2}} \put(172,52){\vector(3,-2){2}}
\put(172,78){\vector(3,2){2}} \put(190,68){\vector(0,-2){2}}
\put(206,89){\vector(3,0){2}} \put(206,41){\vector(3,0){2}}

\put(126,70){$D$} \put(164,38){$K^*$} \put(164,83){$K^*$}
\put(194,62){$\pi$} \put(216,85){$K$} \put(216,36){$K$}
\put(180,5){(f)}


\put(246,65){\line(1,0){24}} \put(271,65){\circle*{6}}
\put(273,65){\line(3,2) {33}} \put(273,65){\line(3,-2){33}}
\put(309,42){\circle*{6}} \put(309,88){\circle*{6}}
\put(310,44){\line(0,1){44}} \put(310,41){\line(1,0){24}}
\put(310,89){\line(1,0){24}}

\put(258,65){\vector(3,0){2}} \put(292,52){\vector(3,-2){2}}
\put(292,78){\vector(3,2){2}} \put(310,68){\vector(0,-2){2}}
\put(326,89){\vector(3,0){2}} \put(326,41){\vector(3,0){2}}

\put(246,70){$D$} \put(285,38){$K^*$} \put(285,83){$K^*$}
\put(314,62){$\eta$} \put(336,85){$K$} \put(336,36){$K$}
\put(300,5){(g)}

\put(366,65){\line(1,0){24}} \put(391,65){\circle*{6}}
\put(393,65){\line(3,2) {33}} \put(393,65){\line(3,-2){33}}
\put(429,42){\circle*{6}} \put(429,88){\circle*{6}}
\put(430,44){\line(0,1){44}} \put(430,41){\line(1,0){24}}
\put(430,89){\line(1,0){24}}

\put(378,65){\vector(3,0){2}} \put(412,52){\vector(3,-2){2}}
\put(412,78){\vector(3,2){2}} \put(430,68){\vector(0,-2){2}}
\put(446,89){\vector(3,0){2}} \put(446,41){\vector(3,0){2}}

\put(366,70){$D$} \put(407,41){$\rho$} \put(407,83){$\phi$}
\put(434,62){$K$} \put(456,85){$K$} \put(456,36){$K$}
\put(420,5){(h)}
\end{picture}
\end{center}

\caption{{\small Intermediate states in rescattering process for
$D\to KK$ decays}} \label{scatter3}
\end{figure}

In order to calculate FSI contribution of $D$ decays to $\pi\pi$,
$K\pi$ and $KK$ one needs to analyze which channel can rescatter
into the final states. The rescattering processes are
$D\to\pi\pi\to\pi\pi$, $D\to KK\to\pi\pi$,
$D\to\rho\rho\to\pi\pi$, $D\to K^*K^*\to\pi\pi$ for $D\to\pi\pi$
decays; $D\to K\pi \to K\pi$, $D\to K^*\rho \to K\pi$ for $D\to
K\pi$ channels; $D\to\pi\pi\to KK$, $D\to KK\to KK$, $D\to
\pi\eta\to KK$, $D\to\rho\rho\to KK$, $D\to K^*K^*\to KK$,
$D\to\rho\phi\to KK$ for $D\to KK$ decays and pictorially shown in
Fig. \ref{scatter1}, Fig. \ref{scatter2} and Fig. \ref{scatter3}.
These rescattering processes give the largest contributions,
because the intermediate states have the largest couplings with
the final states and the masses of exchanged meson are small which
give the largest $t$-channel contributions. When we calculate the
contribution of each diagram in Figs. \ref{scatter1} $\sim$
\ref{scatter3} via eqs.(\ref{fsitv}) and (\ref{fsitp}), we should,
at first, consider all the possible isospin structure for each
diagram and draw all the possible sub-diagrams on the quark level.
Secondly, write down the isospin factor for each sub-diagram. For
example, the $u\bar{u}$ component in one final meson $\pi^0$
contributes an isospin factor $\frac{1}{\sqrt{2}}$, and the
$d\bar{d}$ component contributes $-\frac{1}{\sqrt{2}}$. For the
intermediate state $\pi^0$, the factor $\frac{1}{\sqrt{2}}$ and
$-\frac{1}{\sqrt{2}}$ should be dropped, otherwise, the isospin
relation between different channels would be violated~\cite{ady}.
Third, sum the contributions of all the possible sub-diagrams on
the quark level to get the isospin factor for each diagram on the
hadronic level.

The FSI contributions of the Cabibbo suppressed decays $D^+\to
\pi^+\pi^0$ and $D^0\to \pi^0\pi^0$ depend on the couplings and
phases $g_{K^*K\pi}e^{i\theta_{K^*K\pi}}$ and
$g_{\rho\pi\pi}e^{i\theta\rho\pi\pi}$respectively, while $D^0\to
\pi^+\pi^-$ depends on both of them. In the Cabibbo favored decays
$D^+\to \bar K^0\pi^+$, $D^0\to K^-\pi^+$, $D^0\to \bar K^0\pi^0$
and the doubly Cabibbo suppressed decays $D^0\to K^+\pi^-$,
$D^+\to K^+\pi^0$, $D^+\to K^0\pi^+$ and $D^0\to K^0\pi^0$, the
branching ratios, including both the direct decay and the
rescattering effect, depend not only on
$g_{K^*K\pi}e^{i\theta_{K^*K\pi}}$ and
$g_{\rho\pi\pi}e^{i\theta\rho\pi\pi}$ but also $g_{\rho
KK}e^{i\theta\rho KK}$. For the Cabibbo suppressed decay modes
$D^+\to K^+\bar K^0$, $D^0\to K^+K^-$ and $D^0\to K^0\bar K^0$,
the FSI effects come from $g_{K^*K\pi}e^{i\theta_{K^*K\pi}}$,
$g_{\rho KK}e^{i\theta\rho KK}$, $g_{\phi KK}e^{i\theta_{\phi
KK}}$ and $g_{K^*K\eta}e^{i\theta_{K^*K\eta}}$.

In the numerical calculation, we use the input parameters: 1) the
decay constants, $f_{\pi}=0.133$ GeV, $f_{K}=0.162$ GeV,
$f_{\rho}=0.2$ GeV, $f_{K^*}=0.221$ GeV, $f_{\phi}=0.233$ GeV; 2)
the form factors, $F^{D\pi}(0)= 0.692$, $F^{DK}(0)= 0.762$,
$A_1^{DK^*}(0)=0.880$, $A_2^{DK^*}(0)=1.147$,
$A_1^{D\rho}(0)=0.775$, $A_2^{D\rho}(0)=0.923$~\cite{BSW}. Except
for the decay constants, the values of the form factors have not
been known exactly yet. We therefore have to take them from
model-dependent calculations. The parameter $\Lambda$ in the
off-shellness compensating function $F(k^2)$ introduced in
eq.(\ref{fsisk}) takes the value of $0.513 GeV$, which is in the
mass ranges of the final state mesons. In order to get the
branching ratios which include both the direct decays and the
rescattering effects, we use  eq.(\ref{couple}) and the center
values of the measured decay width of $K^*\to K\pi$,
$\rho\to\pi\pi$ and $\phi\to KK$~\cite{PDG} to obtain
$g_{K^*K\pi}=4.59$, $g_{\rho \pi\pi}=6.0$ and $g_{\phi KK}=5.77$.
We take $g_{\rho KK}=\sqrt{\lambda}g_{\rho \pi\pi}$ with the
$s\bar s$ suppression parameter $\lambda$=0.28~\cite{pk}. Since
there is no data existing for $K^*\to K\eta$ (the mass of $K^*$ is
not large enough to decay into $K\eta$), we estimate the value of
the strong coupling $g_{K^*K\eta}\simeq 3.5$ by comparing
$g_{\rho\pi\pi}$ with $g_{K^*K\pi}$, and considering $SU(3)$
flavor symmetry with $20\%\sim 30\%$ violation. When the
nonfactorizable parameter $\chi$ is not taken into account, we can
not reproduce the experimental data for all the $D\to PP$ decays
simultaneously. So we need to keep it as a phenomenological
parameter. By taking $\chi=-8.6\times 10^{-2}$, the experimental
data of all the detected $D\to PP$ decays can be well accommodated
within the experimental errors.

The strong phases of the effective hadronic couplings
$\theta_{K^*K\pi}$, $\theta_{\rho\pi\pi}$, $\theta_{\rho KK}$,
$\theta_{\phi KK}$ and $\theta_{K^*K\eta}$ can not be known from
direct experimental measurement or from nonperturbative
calculations, because there are no any such kind of computations
yet. The only information is that the values of these phases
should not differ too much, according to $SU(3)$ flavor symmetry.
We fit the experimental data to get the values for these phase
parameters, and find that it is possible to reproduce the
experimental data of these $D$ decays with small $SU(3)$ flavor
symmetry violation. To show this situation, in which the
experimental data are accommodated, Table \ref{tabbr2} gives the
numerical results of the branching ratios at
$\theta_{K^*K\pi}=53.9^\circ$, $\theta_{\rho\pi\pi}=57.3^\circ$,
$\theta_{\rho KK}=71.8^\circ$, $\theta_{\phi KK}=58.7$ and
$\theta_{K^*K\eta}=65^\circ$, with a small $SU(3)$ symmetry
breaking effects. Column `Factorization' is for the branching
ratio predicted in naive factorization approach, where the
nonfactorizable correction is small. We find that the data of $D
\to PP$ cannot be accommodated without including the contribution
of the nonfactorizable effect. Column `Factorization + FSI' is for
the branching ratio of naive factorization including the final
state interaction. The contributions of final state rescattering
effects are large, which can improve the predictions of naive
factorization to be consistent with the experimental data. The
strong phases introduced for the effective hadronic couplings
$g_{K^*K\pi}$, $g_{\rho\pi\pi}$, $g_{\rho KK}$, $g_{\phi KK}$ and
$g_{K^*K\eta}$ are important for explaining the experimental data,
otherwise, it is quite difficult to get the correct results for
these decay modes at the same time by varying other input
parameters.

\begin{table}
\caption{{\small The branching ratios of $D \to PP$.}}
\begin{center}
\begin{tabular}{|c|c|c|c|} \hline
Decay mode & Factorization & Factorization + FSI & Experiment\\
\hline $ D^+ \to \pi^+\pi^0$ &$2.71 \times 10^{-3}$&
$1.8\times 10^{-3}$&$(2.5\pm 0.7)\times 10^{-3}$ \\

$D^0 \to \pi^+\pi^-$ &$2.65 \times 10^{-3}$&
$1.49\times 10^{-3}$ &$(1.43 \pm 0.07) \times 10^{-3}$\\

$D^0 \to \pi^0\pi^0$ &$1.39 \times 10^{-5}$ &
$1.06\times 10^{-3}$& $(8.4 \pm 2.2) \times 10^{-4}$\\

$D^+ \to \bar K^0 \pi^+$ &$9.98 \times 10^{-2}$ &
$2.95\times 10^{-2}$& $(2.77 \pm 0.18) \times 10^{-2}$\\

$D^0 \to K^- \pi^+$ &$5.14 \times 10^{-2}$ &
$3.72\times 10^{-2}$& $(3.80 \pm 0.09) \times 10^{-2}$\\

$D^0 \to \bar K^0 \pi^0$ &$4.1 \times 10^{-4}$ &
$2.09\times 10^{-2}$& $(2.28 \pm 0.22) \times 10^{-2}$\\

$D^0 \to K^+ \pi^-$ &$2.01 \times 10^{-4}$ &
$1.41\times 10^{-4}$& $(1.48 \pm 0.21) \times 10^{-4}$\\

$D^+ \to K^+ \pi^0$ &$2.56 \times 10^{-4}$ &
$2.96\times 10^{-4}$& -\\

$D^+ \to K^0 \pi^+$ &$5.39 \times 10^{-6}$ &
$7.56\times 10^{-4}$& -\\

$D^0 \to K^0 \pi^0$ &$1.06 \times 10^{-6}$ &
$2.84\times 10^{-4}$& - \\

$D^+ \to K^+\bar K^0$ &$9.76 \times 10^{-3}$ &
$6.4\times 10^{-3}$& $(5.8 \pm 0.6) \times 10^{-3}$\\

$D^0 \to K^+K^-$ &$3.83 \times 10^{-3}$ &
$4.0\times 10^{-3}$& $(4.12 \pm 0.14) \times 10^{-3}$\\

$D^0 \to K^0\bar{K^0}$ &$0$ &
$5.73\times 10^{-4}$& $(7.1 \pm 1.9) \times 10^{-4}$\\
\hline
\end{tabular}
\end{center}
\label{tabbr2}
\end{table}
As we have mentioned earlier, the branching ratios have not been
detected in experiment for doubly Cabibbo-suppressed decay modes
$D^+ \to K^+ \pi^0$, $D^+ \to K^0 \pi^+$ and $D^0 \to K^0 \pi^0$,
except $D^0 \to K^+ \pi^-$ yet. In our method, they are all
predicted to be at order of $O(10^{-4})$.

To conclude this section, we shall  give some comments. There are
some free parameters, such as the $D$ decay form factors which
have not been well determined in experiment yet. They need to be
measured from leptonic and semileptonic decays of $D$ mesons,
which are quite possible in CLEO-C program in the near future. The
other input parameters that may cause uncertainties are the shape
of the off-shell compensating function $F(k^2)$ and the
nonfactorizable parameter $\chi$, which are needed to be studied
by some nonperturbative methods based on QCD in the future.
Certainly to completely  understand final state interactions, more
experimental data and more theoretical works are needed.

{\bf Summary} We have studied two-body nonleptonic charmed meson
decays into two pseudoscalar mesons.  The total decay amplitude
includes both direct weak decays and final state rescattering
effects. The direct weak decays are calculated in factorization
approach, and the final state interaction effects are studied in
one-particle-exchange method. The prediction of naive
factorization is far from the experimental data. After including
the contribution of final state interaction, as well as the
nonfactorizable correction, the theoretical predictions can
accommodate the experimental data within experimental errors,
where the strong phases of the effective couplings are quite
necessary to reproduce experimental data. The branching ratios are
predicted for the three doubly Cabibbo-suppressed decay modes.

\vspace{5mm}

\noindent {\large{\bf Acknowledgement}}\vspace{0.3cm}

\noindent This work is supported in part by National Natural
Science Foundation of China with contract No.10205017,
and by the Grant of BEPC National Laboratory.
M. Ablikim is grateful to Scientific
Research Foundation for  Returned Scholars of State Education
Ministry of China.


\begin{thebibliography}{9}

\bibitem{BSW}   M. Wirbel, B. Stech and M. Bauer, Z. Phys. \textbf{C29}, 637 (1985);
               M. Bauer, B. Stech and M. Wirbel, Z. Phys. \textbf{C34}, 103 (1987).

\bibitem{fsi} H.J. Lipkin, Phys. Rev. Lett. \textbf{44}, 710 (1980);
              J.F. Donoghue and B.R. Holstein, Phys. Rev. \textbf{D21}, 1334 (1980).

\bibitem{Donoghue}J.F. Donoghue, Phys. Rev. \textbf{D33}, 1516 (1986).

\bibitem{ope1}Y. Lu, B.S. Zou and M.P. Locher,  Z. Phys. \textbf{A345}, 207 (1993);
             M.P. Locher, Y. Lu and B.S. Zou, Z. Phys. \textbf{A347}, 281 (1994);
             Y. Lu and M.P. Locher, Z. Phys. \textbf{A351}, 83 (1995).

\bibitem{ope2}X.Q. Li and B.S. Zou, Phys. Lett. \textbf{B399}, 297 (1997);
             Y.S. Dai, D.S. Du, X.Q. Li, Z.T. Wei and B.S. Zou, Phys. Rev.
             \textbf{D60}, 014014 (1999).

\bibitem{ref} A.N. Kamal and R.C. Verma, Phys. Rev. \textbf{D35}, 3515 (1987);
              A.N. Kamal and R. Sinha, Phys. Rev. \textbf{D36}, 3510 (1987);
         H.J. Lipkin, Phys. Lett. \textbf{B 283}, 421 (1992);
         T.N. Pham, Phys. Rev. \textbf{D46}, 2976 (1992);
         L.L. Chau and H.Y. Cheng, Phys. Lett. \textbf{B333}, 514 (1994);
       X.Q. Li and B.S. Zou, Phys. Rev. \textbf{D57}, 1518 (1998).

\bibitem{r} J.L. Rosner, Phys. Rev. \textbf{D60}, 114026 (1999);
           C.W. Chiang and J.L. Rosner, Phys. Rev. \textbf{D65}, 054007 (2002);
           C.W. Chiang, Z. Luo and J.L. Rosner, hep-ph/0209272.

\bibitem{cheng} H.Y. Cheng, hep-ph/0202254.

\bibitem{qd} L.L. Chau and H.Y. Cheng, Phys. Rev. \textbf{D36}, 137 (1987); Phys. Lett.
              \textbf{B222}, 285 (1989); M. Gronau, O.F. Hern\'andez, D. London
              and J.L. Rosner, Phys. Rev. \textbf{D50}, 4529 (1994); Phys. Rev. \textbf{D52}, 6356 (1995).

\bibitem{ady} M. Ablikim, D.S. Dong and M.Z. Yang, Phys. Lett.
              \textbf{B536}, 34 (2002).

\bibitem{ldy} J.W. Li, M.Z. Yang and D.S. Dong, hep-ph/0206154.

\bibitem{buras} For a review see, G. Buchalla, A.J. Buras and M.E. Lautenbacher,
        Rev. Mod. Phys. \textbf{68}, 1125 (1996).

\bibitem{BBNS} M. Beneke, G. Buchalla, M. Neubert and C.T. Sachrajda,
        Phys. Rev. Lett. \textbf{83}, 1914 (1999); Nucl. Phys. \textbf{B591}, 313 (2000);
        Nucl. Phys. \textbf{B606}, 245 (2001).

\bibitem{PDG} Particle Data Group, Phys. Rev. \textbf{D66}, 010001 (2002).

\bibitem{Fk} O. Gortchakov, M.P. Locher, V.E. Markushin and S. von Rotz,
      Z. Phys. \textbf{A353}, 447 (1996).

\bibitem{pk} K. Peters and E. Klempt, Phys. Lett. \textbf{B352}, 467 (1995).


\end{thebibliography}
\end{document}